\documentclass[traditabstract]{aa}
\usepackage{graphicx}
\usepackage{txfonts}

\usepackage{natbib}
\usepackage{color}
\usepackage{overpic}
\begin{document} 

   \title{Excitation and charge transfer in low-energy hydrogen atom collisions with neutral oxygen \thanks{Data available in electronic form at the CDS via anonymous ftp to cdsarc.u-strasbg.fr (130.79.128.5) 
   or via http://cdsweb.u-strasbg.fr/cgi-bin/qcat?J/A+A/.  The data are also available at https://github.com/barklem/public-data.}}
   
   \titlerunning{Low-energy O+H collisions}


\author{P. S. Barklem\inst{1}
          }

   \institute{Theoretical Astrophysics, Department of Physics and Astronomy, Uppsala University,
              Box 516, SE-751 20 Uppsala, Sweden}

   \date{Received 19 September 2017 ; accepted 22 November 2017}

 
  \abstract
   {Excitation and charge transfer in low-energy O+H collisions is studied; it is a problem of importance for modelling stellar spectra and obtaining accurate oxygen abundances in late-type stars including the Sun.  The collisions have been studied theoretically using a previously presented method based on an asymptotic two-electron linear combination of atomic orbitals (LCAO) model of ionic-covalent interactions in the neutral atom-hydrogen-atom system, together with the multichannel Landau-Zener model.  The method has been extended to include configurations involving excited states of hydrogen using an estimate for the two-electron transition coupling, but this extension was found to not lead to any remarkably high rates.  Rate coefficients are calculated for temperatures in the range 1000--20000~K, and charge transfer and (de)excitation processes involving the first excited $S$-states, $4s.^5$S$^o$ and $4s.^3$S$^o$, are found to have the highest rates.}


   \maketitle
%
\section{\label{sec:intro}Introduction}

As the third most abundant element in the universe, oxygen is of great importance.  Oxygen has a large impact on stellar structure, and thus the precise abundance of oxygen in the Sun lies at the heart of the current `solar oxygen problem', which  is the inability to reconcile theoretical solar models adopting the surface oxygen abundance derived from spectroscopy with helioseismic observations \citep{bahcall_what_2004, bahcall_new_2005, delahaye_solar_2006, basu_helioseismology_2008, serenelli_new_2009, basu_understanding_2015}.   No simple solution has been found and it is clear that progress must be made on various aspects of the solar models, not least the interior opacities \citep{bailey_higher-than-predicted_2015, mendoza_computation_2017}, and  on the accuracy and reliability of the measurement of the solar oxygen abundance \citep{asplund_line_2004,steffen_photospheric_2015}.  Oxygen is also an important element in understanding Galactic chemical evolution \citep[see e.g.][]{stasinska_oxygen_2012,amarsi_galactic_2015}, and in understanding the chemistry of stellar systems with exoplanets \citep{madhusudhan_c/o_2012, nissen_carbon_2014}.

The only strong feature of oxygen available in late-type stars is the \ion{O}{i} triplet lines at 777~nm, which from comparisons of models with centre-to-limb observations on the Sun is well known to form out of local thermodynamic equilibrium (LTE) \citep{altrock_new_1968, eriksson_o_1979, kiselman_777_1993, AllendePrieto2004b, Pereira2009, Pereira2009a}.  Modelling without the assumption of LTE,  in statistical equilibrium, requires that the effects of all radiative and collisional processes on the atom of interest  be known.  Collisional data with electrons and hydrogen atoms are of most relevance in late-type stars \citep[][and references therein]{Lambert1993, barklem_accurate_2016}.  Data for electron collisions on \ion{O}{i} are reasonably well established, with $R$-matrix calculations available that are in good agreement \citep{zatsarinny_r-matrix_2002, Barklem2007a, tayal_b_2016}.  However, for hydrogen collisions basically all astrophysical modelling has either relied on simple classical estimates or omitted these processes.   In particular, the Drawin formula, a modified classical Thomson model approach \citep{Drawin1968, Drawin1969, Drawin1973, steenbock_statistical_1984}, usually with a scaling factor $S_H$ calibrated on the centre-to-limb observations in the Sun has been used \citep[e.g.][]{steffen_photospheric_2015}.  Some studies \citep[e.g.][]{asplund_line_2004} have preferred to omit these processes altogether on the basis that the Drawin formula compares poorly to existing experiment and detailed calculations.

Thus, to put oxygen abundances in late-type stars, including the Sun, on a firm footing, requires reliable data for O+H collision processes (see e.g. \cite{Asplund2005} and  \cite{barklem_accurate_2016} for reviews).  Previous work on inelastic hydrogen collision processes with neutral atoms, for the one available experiment \citep{Fleck1991} and for the detailed calculations for simple atoms \citep{Belyaev1999, Belyaev2003, Belyaev2010, Guitou2011, Belyaev2012} has shown the importance of the ionic crossing mechanism, leading to both excitation and charge transfer processes.  Such detailed calculations are time consuming and difficult for systems involving complex atoms, and so in order to be able to obtain estimates of the processes with the highest rates for the many atoms needed in astrophysical modelling including complex atoms, asymptotic model approaches considering the ionic crossing mechanism have recently been put forward.  In particular, a semi-empirical approach has been developed \citep{Belyaev2013} based on a fitting formula to the coupling based on measured and calculated values \citep{Olson1971} and applied in a number of studies \citep{Belyaev2014, belyaev_model_2016, belyaev_atomic_2017}.  In addition, a theoretical two-electron linear combination  of atomic orbitals (LCAO) method has been developed \citep{barklem_excitation_2016, barklem_erratum:_2017}, based on earlier work \citep{Grice1974, Adelman1977, Anstee1992}.  Comparisons with detailed calculations for the simple atoms show that both methods perform well in identifying the processes with the highest rates and in estimating these rates to order-of-magnitude accuracy.  Astrophysical modelling of simple atoms has shown that this is sufficient in such cases \citep{Barklem2003b, Lind2011, osorio_mg_2015}.  Whether this situation can be extrapolated to complex atoms is unclear; however, calculating estimates of rates for the complex systems based on these model approaches provides a much  sounder basis for progress than the currently employed classical estimates.  In the present paper, calculations for O+H using the LCAO model are presented.  Calculations with the semi-empirical estimate of the coupling \citep{Olson1971}, as well as Landau-Herring method estimates \citep{smirnov_formation_1965, smirnov_negativeion_1967, Janev1976}, are done in order to investigate the sensitivity of the results to the coupling, and thus obtain some indication of the uncertainty due to this source.

\section{Calculations}

Calculations have been performed using the method and codes described in \cite{barklem_excitation_2016, barklem_erratum:_2017} (hereafter B16) and the notation here follows that paper.  Calculations are done for potentials and couplings from the LCAO method described in that paper, and for the semi-empirical formula of \cite{Olson1971}, and the Landau-Herring method as derived by \cite{Janev1976} and \cite{smirnov_formation_1965, smirnov_negativeion_1967}.  Hereafter, these models are  referred to as LCAO, SEMI-EMP, LH-J, and LH-S, respectively.  Many aspects of the codes have been improved, including the ability to handle covalent states in which hydrogen is excited to the $n=2$ state.  This is necessary for O+H, as the comparable ionisation energies of O and H means that covalent states dissociating to $\mathrm{O}(2p^4.^3P) + \mathrm{H}(n=2)$ are below the ionic limit.  This could lead to processes such as
\begin{equation}
\mathrm{O}(2p^4.^3P) + \mathrm{H}(n=2) \rightleftarrows \mathrm{O}^* + \mathrm{H}(1s),
\end{equation}
potentially with small thresholds.  In the current model, such a process proceeds via interaction of the $\mathrm{O}(2p^4.^3P) + \mathrm{H}(n=2)$ covalent state with an ionic state $\mathrm{O}^+(2p^3) + \mathrm{H}^-$.  Such a non-adiabatic transition corresponds to a two-electron process, and the appropriate coupling has been calculated using the expression presented by \cite{belyaev_atomic_2017}, which is based on work in \cite{Belyaev1993}, an estimate that expresses the coupling for a two-electron transition $H_{1j}^{2e}(R)$ in terms of the corresponding coupling for the one-electron transition case $H_{1j}^{1e}(R)$, namely
\begin{equation}
H_{1j}^{2e}(R) = \left\{H_{1j}^{1e}(R) \right\}^2 \times R,
\end{equation}
where all quantities are in atomic units.  We note that the LCAO model is called a two-electron model since it describes two-electrons explicitly, but gives the coupling between states corresponding to a one-electron transition.  If the interaction of an ionic state $\mathrm{O}^+(2p^3) + \mathrm{H}^-$ with the covalent state $\mathrm{O}(2p^4.^3P) + \mathrm{H}(n=2)$ is considered, the interpretation of this process following from \cite{Belyaev1993} is that the two electrons on H$^-$ simultaneously transfer towards the O$^+$ core, but due to the lack of a possible state accepting both electrons, one electron ends up in an excited state on the proton.

A few other small changes have been made in the codes that are worth noting.  In B16, angular momentum coupling factors $C$ are applied as described in Eqs.~18, 19, and 20 of that paper.  As the SEMI-EMP model formula for the coupling is based on cases where angular momentum coupling plays a role, and not only on cases with spherical symmetry where $C=1$, it is debatable that this factor should be employed in this case.  In order to be consistent with other work \citep{Belyaev2013, Belyaev2014, belyaev_model_2016, belyaev_atomic_2017}, this factor is now always set to $C=1$ for the SEMI-EMP model.  Further, it is noted that to employ eqn.~18 for the two LH models, an extra factor $\sqrt{2}$ is required to account for the two equivalent electrons on H$^-$. In the LCAO model this is not present since both electrons are considered explicitly in the two-electron wavefunction, while the LH model only considers one electron (see e.g. \cite{Chibisov1988} for details).

The considered states and their relevant data used as input for the calculation are presented in Table~\ref{tab:input}. The resulting possible symmetries are given in Table~\ref{tab:syms}, including the five symmetries in which the considered ionic states may occur, and thus which are calculated.  We note that the most excited ionic state $\mathrm{O}^+(2p^3.^2P^o)+\mathrm{H}^-$ is included in the calculations, but only has crossings with the considered covalent states at very short internuclear distance where the asymptotic methods used here are not valid.  The calculations are performed for $R>5$~$a_0$, and thus no transitions relating to this core arise in the calculations, and this core state could be removed.  It was retained, however, for completeness so that all possible parents of the oxygen ground term are explicit.  Four states, $4d.^5D^o$, $4d.^3D^o$, $4f.^5F$, and $4f.^3F$, fall below the first ionic limit corresponding to $\mathrm{O}^+(2p^3.^4S^o) + \mathrm{H}^-$; however, they have not been included in the calculations as the crossings occur at $R>250$~$a_0$, and are practically diabatic.   The crossing with the highest covalent state occurs at around 163~$a_0$, and thus the potential calculations are performed for internuclear distances $R$ ranging from 5 to 200~$a_0$.  Collision dynamics are calculated in the multichannel Landau-Zener model, and the cross sections are calculated for collision energies from thresholds up to 100~eV. The rate coefficients are then calculated for temperatures in the range 1000--20000~K, with steps of 1000~K, for the various models.

\begin{table*}
\center
\caption{\label{tab:input}  Input data for the calculations. The notation from the LCAO model in B16 is used, and detailed descriptions are given in that paper.  In short, $L_A$ and $S_A$ are the electronic orbital angular momentum and spin quantum numbers for the state of the oxygen atom; $n$ and $l$ are the principal and angular momentum quantum numbers for the active electron.  $E_j^\mathrm{O/O^+}$ is the state energy for the oxygen atom, $E_\mathrm{lim}$ the corresponding series limit, and $E_j$ the total asymtotic molecular energy.  The zero point in the case of energies on the oxygen atom, $E_j^\mathrm{O/O^+}$ and $E_\mathrm{lim}$, is the \ion{O}{i} ground term, and the zero point for the asymptotic molecular energies $E_j$ is the energy corresponding to both atoms in their ground states.  $N_\mathrm{eq}$ is the number of equivalent active electrons. $L_C$ and $S_C$ are the electronic orbital angular momentum and spin quantum numbers for the core of the oxygen atom.  $G^{S_A L_A}_{S_c L_c}$ is the coefficient of fractional parentage.   For covalent configurations where oxygen is neutral and/or hydrogen is in the ground state, H($1s$) is implied and omitted for clarity. }
\begin{tabular}{lccccrrrclccr}
\hline \hline
$         \mathrm{Configuration.Term}$ & $         L_A$ & $      2S_A+1$ & $           n$ & $           l$ & $           E_j^\mathrm{O/O^+}$ & $     E_\mathrm{lim}$ & $     E_\mathrm{j}$ & $      N_\mathrm{eq}$ & $          \mathrm{Core}$ & $         L_c$ & $      2S_c+1$ & $    G^{S_A L_A}_{S_c L_c}$ \\ 
  &  &  &  &  & [cm$^{-1}$] & [cm$^{-1}$] & [cm$^{-1}$] &  &  &  &  &  \\ \hline
  &  &  &  &  &  &  &  &  &  &  &  &  \\
\multicolumn{12}{c}{\underline{Covalent states}}  \\
$                                2p^4.^3P$ & $           1$ & $           3$ & $           2$ & $           1$ & $           0$ & $      109759$ & $           0$ & $           4$ & $                        \mathrm{O}^+(^4S^o)$  &$           0$ & $           4$ & $                   -0.577$ \\
$                                2p^4.^3P$ & $           1$ & $           3$ & $           2$ & $           1$ & $           0$ & $      136577$ & $           0$ & $           4$ & $                        \mathrm{O}^+(^2D^o)$  &$           2$ & $           2$ & $                    0.645$ \\
$                                2p^4.^3P$ & $           1$ & $           3$ & $           2$ & $           1$ & $           0$ & $      150228$ & $           0$ & $           4$ & $                        \mathrm{O}^+(^2P^o)$  &$           1$ & $           2$ & $                   -0.500$ \\
$                                2p^4.^1D$ & $           2$ & $           1$ & $           2$ & $           1$ & $       15790$ & $      136577$ & $       15790$ & $           4$ & $                        \mathrm{O}^+(^2D^o)$  &$           2$ & $           2$ & $                    0.866$ \\
$                                2p^4.^1D$ & $           2$ & $           1$ & $           2$ & $           1$ & $       15790$ & $      150228$ & $       15790$ & $           4$ & $                        \mathrm{O}^+(^2P^o)$  &$           1$ & $           2$ & $                   -0.500$ \\
$                                2p^4.^1S$ & $           0$ & $           1$ & $           2$ & $           1$ & $       33715$ & $      150228$ & $       33715$ & $           4$ & $                        \mathrm{O}^+(^2P^o)$  &$           1$ & $           2$ & $                    1.000$ \\
$                                3s.^5S^o$ & $           0$ & $           5$ & $           3$ & $           0$ & $       73690$ & $      109759$ & $       73690$ & $           1$ & $                        \mathrm{O}^+(^4S^o)$  &$           0$ & $           4$ & $                    1.000$ \\
$                                3s.^3S^o$ & $           0$ & $           3$ & $           3$ & $           0$ & $       76717$ & $      109759$ & $       76717$ & $           1$ & $                        \mathrm{O}^+(^4S^o)$  &$           0$ & $           4$ & $                    1.000$ \\
$                2p^4.^3P+\mathrm{H(n=2)}$ & $           1$ & $           3$ & $           2$ & $           1$ & $           0$ & $      109759$ & $       82259$ & $           4$ & $                        \mathrm{O}^+(^4S^o)$  &$           0$ & $           4$ & $                   -0.577$ \\
$                2p^4.^3P+\mathrm{H(n=2)}$ & $           1$ & $           3$ & $           2$ & $           1$ & $           0$ & $      136577$ & $       82259$ & $           4$ & $                        \mathrm{O}^+(^2D^o)$  &$           2$ & $           2$ & $                    0.645$ \\
$                2p^4.^3P+\mathrm{H(n=2)}$ & $           1$ & $           3$ & $           2$ & $           1$ & $           0$ & $      150228$ & $       82259$ & $           4$ & $                        \mathrm{O}^+(^2P^o)$  &$           1$ & $           2$ & $                   -0.500$ \\
$                                  3p.^5P$ & $           1$ & $           5$ & $           3$ & $           1$ & $       86551$ & $      109759$ & $       86551$ & $           1$ & $                        \mathrm{O}^+(^4S^o)$  &$           0$ & $           4$ & $                    1.000$ \\
$                                  3p.^3P$ & $           1$ & $           3$ & $           3$ & $           1$ & $       88553$ & $      109759$ & $       88553$ & $           1$ & $                        \mathrm{O}^+(^4S^o)$  &$           0$ & $           4$ & $                    1.000$ \\
$                                4s.^5S^o$ & $           0$ & $           5$ & $           4$ & $           0$ & $       95399$ & $      109759$ & $       95399$ & $           1$ & $                        \mathrm{O}^+(^4S^o)$  &$           0$ & $           4$ & $                    1.000$ \\
$                                4s.^3S^o$ & $           0$ & $           3$ & $           4$ & $           0$ & $       96147$ & $      109759$ & $       96147$ & $           1$ & $                        \mathrm{O}^+(^4S^o)$  &$           0$ & $           4$ & $                    1.000$ \\
$                                3d.^5D^o$ & $           2$ & $           5$ & $           3$ & $           2$ & $       97343$ & $      109759$ & $       97343$ & $           1$ & $                        \mathrm{O}^+(^4S^o)$  &$           0$ & $           4$ & $                    1.000$ \\
$                                3d.^3D^o$ & $           2$ & $           3$ & $           3$ & $           2$ & $       97411$ & $      109759$ & $       97411$ & $           1$ & $                        \mathrm{O}^+(^4S^o)$  &$           0$ & $           4$ & $                    1.000$ \\
$                                  4p.^5P$ & $           1$ & $           5$ & $           4$ & $           1$ & $       99016$ & $      109759$ & $       99016$ & $           1$ & $                        \mathrm{O}^+(^4S^o)$  &$           0$ & $           4$ & $                    1.000$ \\
$                                  4p.^3P$ & $           1$ & $           3$ & $           4$ & $           1$ & $       99603$ & $      109759$ & $       99603$ & $           1$ & $                        \mathrm{O}^+(^4S^o)$  &$           0$ & $           4$ & $                    1.000$ \\
$                                3s.^3D^o$ & $           2$ & $           3$ & $           3$ & $           0$ & $      101065$ & $      136577$ & $      101065$ & $           1$ & $                        \mathrm{O}^+(^2D^o)$  &$           2$ & $           2$ & $                    1.000$ \\
$                                5s.^5S^o$ & $           0$ & $           5$ & $           5$ & $           0$ & $      102039$ & $      109759$ & $      102039$ & $           1$ & $                        \mathrm{O}^+(^4S^o)$  &$           0$ & $           4$ & $                    1.000$ \\
$                                5s.^3S^o$ & $           0$ & $           3$ & $           5$ & $           0$ & $      102334$ & $      109759$ & $      102334$ & $           1$ & $                        \mathrm{O}^+(^4S^o)$  &$           0$ & $           4$ & $                    1.000$ \\
$                                3s.^1D^o$ & $           2$ & $           1$ & $           3$ & $           0$ & $      102584$ & $      136577$ & $      102584$ & $           1$ & $                        \mathrm{O}^+(^2D^o)$  &$           2$ & $           2$ & $                    1.000$ \\
  &  &  &  &  &  &  &  &  &  &  &  &  \\
  \multicolumn{13}{c}{\underline{Ionic states}}  \\
$        \mathrm{O}^+(^4S^o)+\mathrm{H}^-$ & $           0$ & $           4$ & $           -$ & $           -$ & $      109759$ & $           -$ & $      103676$ & $           -$ & $                                           $  &$            $ & $            $ & $                         $ \\
$        \mathrm{O}^+(^2D^o)+\mathrm{H}^-$ & $           2$ & $           2$ & $           -$ & $           -$ & $      136577$ & $           -$ & $      130494$ & $           -$ & $                                           $  &$            $ & $            $ & $                         $ \\
$        \mathrm{O}^+(^2P^o)+\mathrm{H}^-$ & $           1$ & $           2$ & $           -$ & $           -$ & $      150228$ & $           -$ & $      144145$ & $           -$ & $                                           $  &$            $ & $            $ & $                         $ \\ 
\hline
\end{tabular}
\end{table*}

\begin{table*}
\center
\caption{\label{tab:syms} Possible symmetries for O+H molecular states arising from various asymptotic atomic states, and the total statistical weights.  The symmetries leading to covalent-ionic interactions among the considered states, and thus which need to be calculated, are shown at the bottom along with their statistical weights. For covalent configurations where oxygen is neutral and/or hydrogen is in the ground state, this information is implied and omitted for clarity. }
\begin{tabular}{rlcl}
\hline \hline
Label & Configuration.Term & $g_\mathrm{total}$& Terms\\ \hline
$  1$ & $                                2p^4.^3P$ &   18& $   ^{2}\Sigma^-,\        ^{2}\Pi,\   ^{4}\Sigma^-,\        ^{4}\Pi$ \\
$  2$ & $                                2p^4.^1D$ &   10& $   ^{2}\Sigma^+,\        ^{2}\Pi,\     ^{2}\Delta$ \\
$  3$ & $                                2p^4.^1S$ &    2& $   ^{2}\Sigma^+$ \\
$  4$ & $                                3s.^5S^o$ &   10& $   ^{4}\Sigma^-,\   ^{6}\Sigma^-$ \\
$  5$ & $                                3s.^3S^o$ &    6& $   ^{2}\Sigma^-,\   ^{4}\Sigma^-$ \\
$  6$ & $                2p^4.^3P+\mathrm{H(n=2)}$ &   72& $   ^{2}\Sigma^-,\        ^{2}\Pi,\   ^{4}\Sigma^-,\        ^{4}\Pi,\   ^{2}\Sigma^+,\     ^{2}\Delta,\   ^{4}\Sigma^+,\     ^{4}\Delta$ \\
$  7$ & $                                  3p.^5P$ &   30& $   ^{4}\Sigma^-,\        ^{4}\Pi,\   ^{6}\Sigma^-,\        ^{6}\Pi$ \\
$  8$ & $                                  3p.^3P$ &   18& $   ^{2}\Sigma^-,\        ^{2}\Pi,\   ^{4}\Sigma^-,\        ^{4}\Pi$ \\
$  9$ & $                                4s.^5S^o$ &   10& $   ^{4}\Sigma^-,\   ^{6}\Sigma^-$ \\
$ 10$ & $                                4s.^3S^o$ &    6& $   ^{2}\Sigma^-,\   ^{4}\Sigma^-$ \\
$ 11$ & $                                3d.^5D^o$ &   50& $   ^{4}\Sigma^-,\        ^{4}\Pi,\     ^{4}\Delta,\   ^{6}\Sigma^-,\        ^{6}\Pi,\     ^{6}\Delta$ \\
$ 12$ & $                                3d.^3D^o$ &   30& $   ^{2}\Sigma^-,\        ^{2}\Pi,\     ^{2}\Delta,\   ^{4}\Sigma^-,\        ^{4}\Pi,\     ^{4}\Delta$ \\
$ 13$ & $                                  4p.^5P$ &   30& $   ^{4}\Sigma^-,\        ^{4}\Pi,\   ^{6}\Sigma^-,\        ^{6}\Pi$ \\
$ 14$ & $                                  4p.^3P$ &   18& $   ^{2}\Sigma^-,\        ^{2}\Pi,\   ^{4}\Sigma^-,\        ^{4}\Pi$ \\
$ 15$ & $                                3s.^3D^o$ &   30& $   ^{2}\Sigma^-,\        ^{2}\Pi,\     ^{2}\Delta,\   ^{4}\Sigma^-,\        ^{4}\Pi,\     ^{4}\Delta$ \\
$ 16$ & $                                5s.^5S^o$ &   10& $   ^{4}\Sigma^-,\   ^{6}\Sigma^-$ \\
$ 17$ & $                                5s.^3S^o$ &    6& $   ^{2}\Sigma^-,\   ^{4}\Sigma^-$ \\
$ 18$ & $                                3s.^1D^o$ &   10& $   ^{2}\Sigma^-,\        ^{2}\Pi,\     ^{2}\Delta$ \\
$ 19$ & $        \mathrm{O}^+(^4S^o)+\mathrm{H}^-$ &    4& $   ^{4}\Sigma^-$ \\
$ 20$ & $        \mathrm{O}^+(^2D^o)+\mathrm{H}^-$ &   10& $   ^{2}\Sigma^-,\        ^{2}\Pi,\     ^{2}\Delta$ \\
$ 21$ & $        \mathrm{O}^+(^2P^o)+\mathrm{H}^-$ &    6& $   ^{2}\Sigma^+,\        ^{2}\Pi$ \\
&&&\\ 
     \multicolumn{3}{l}{Number of symmetries to calculate :   5} & $    ^{4}\Sigma^-,\   ^{2}\Sigma^-,\        ^{2}\Pi,\     ^{2}\Delta,\   ^{2}\Sigma^+$ \\
 \multicolumn{3}{l}{$g_\mathrm{total}:$} & $              4,\              2,\              4,\              4,\              2$ \\
\hline 
\end{tabular}
\end{table*}

\section{Results and discussion}

\begin{figure*}
\centering
\begin{overpic}[width=0.48\textwidth]{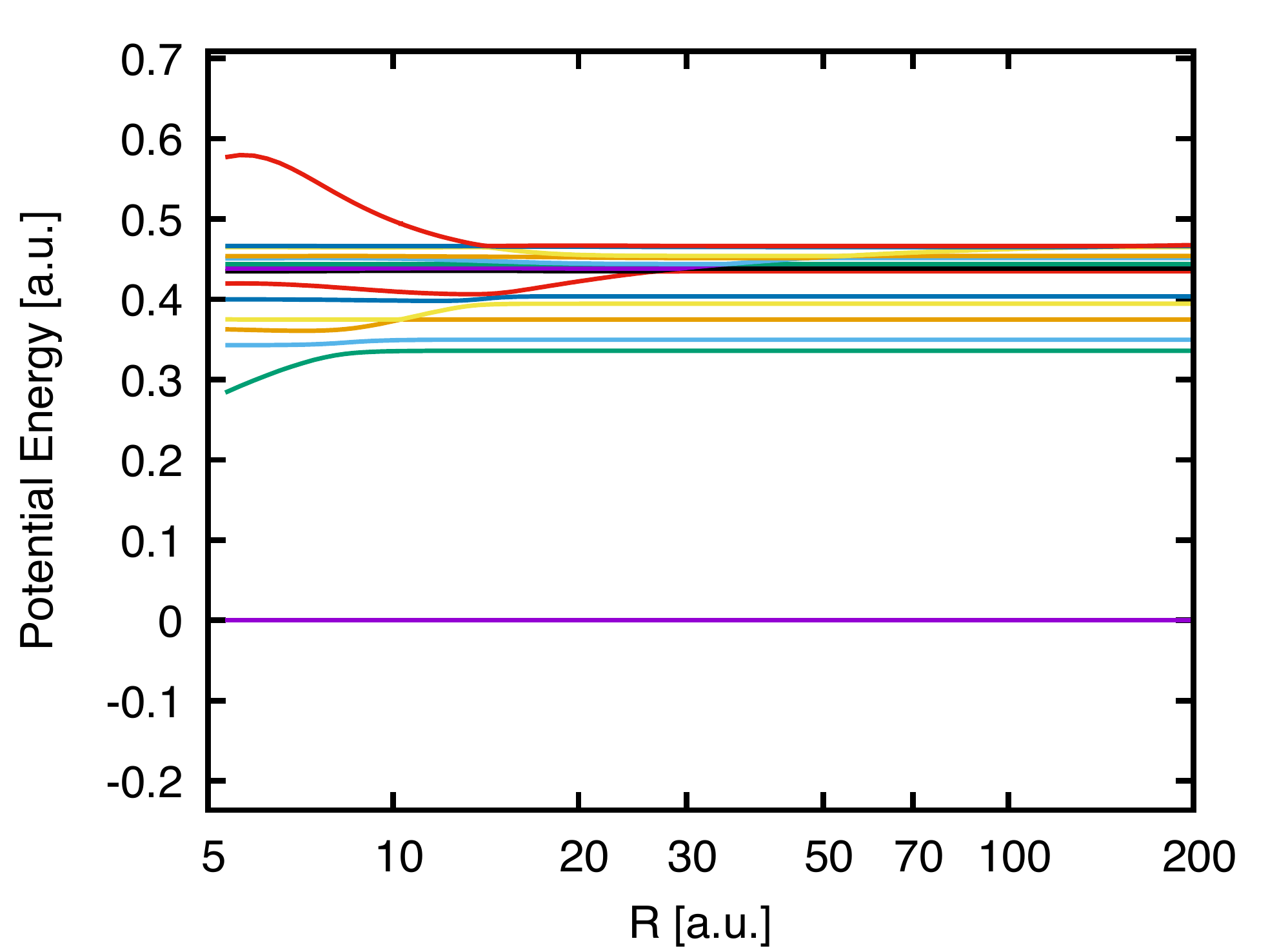}\put(25,17){$^4\Sigma^-$, $\mathrm{O}^+(^4S^o)$ core}
\put(85,27){1}
\put(85,44){4}
\put(85,57){19}
\end{overpic}
\begin{overpic}[width=0.48\textwidth]{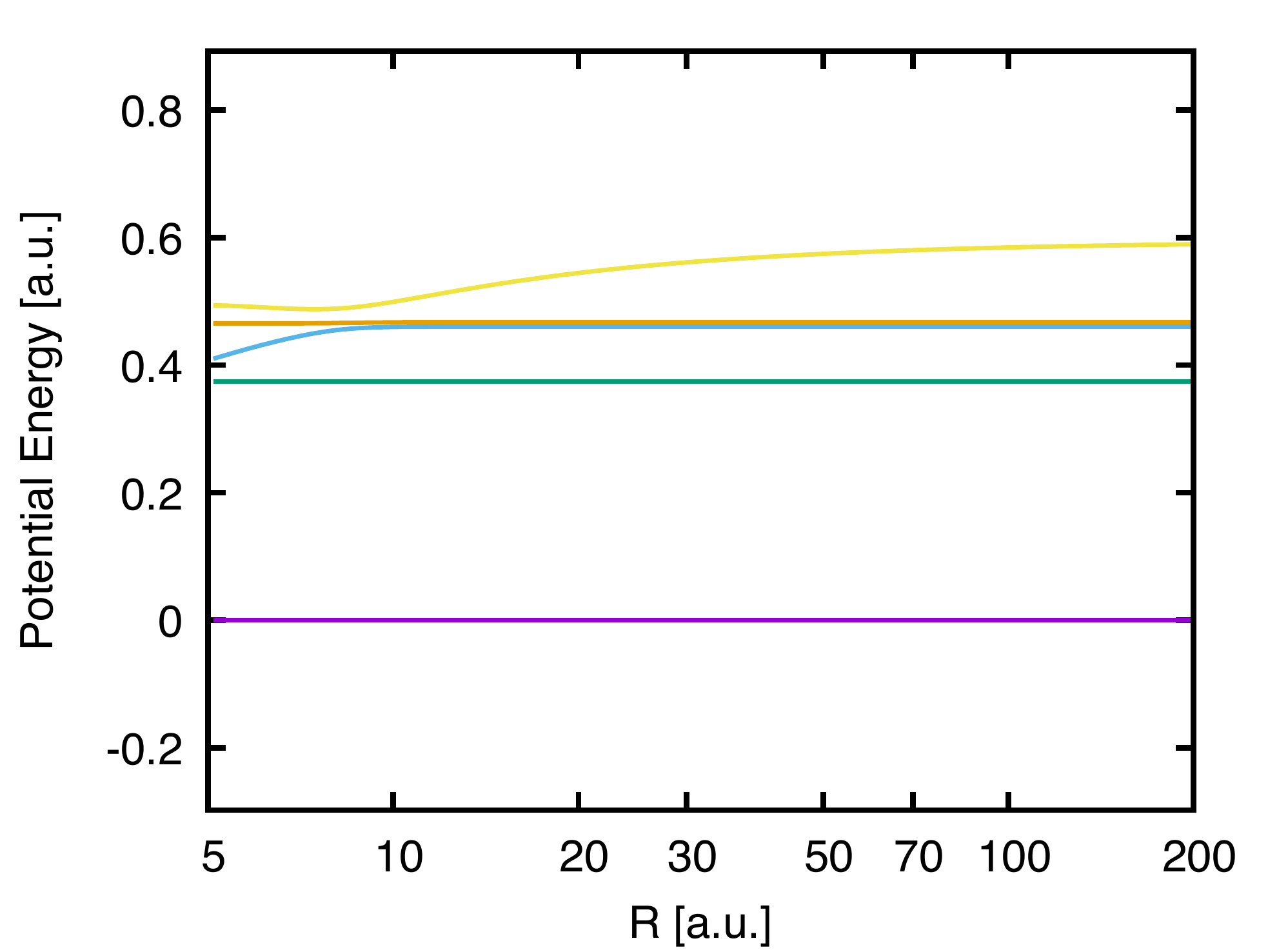}\put(25,17){$^2\Sigma^-$, $\mathrm{O}^+(^2D^o)$ core}
\put(85,27){1}
\put(85,42){6}
\put(85,46){15}
\put(85,50){18}
\put(85,57){20}
\end{overpic}
\begin{overpic}[width=0.48\textwidth]{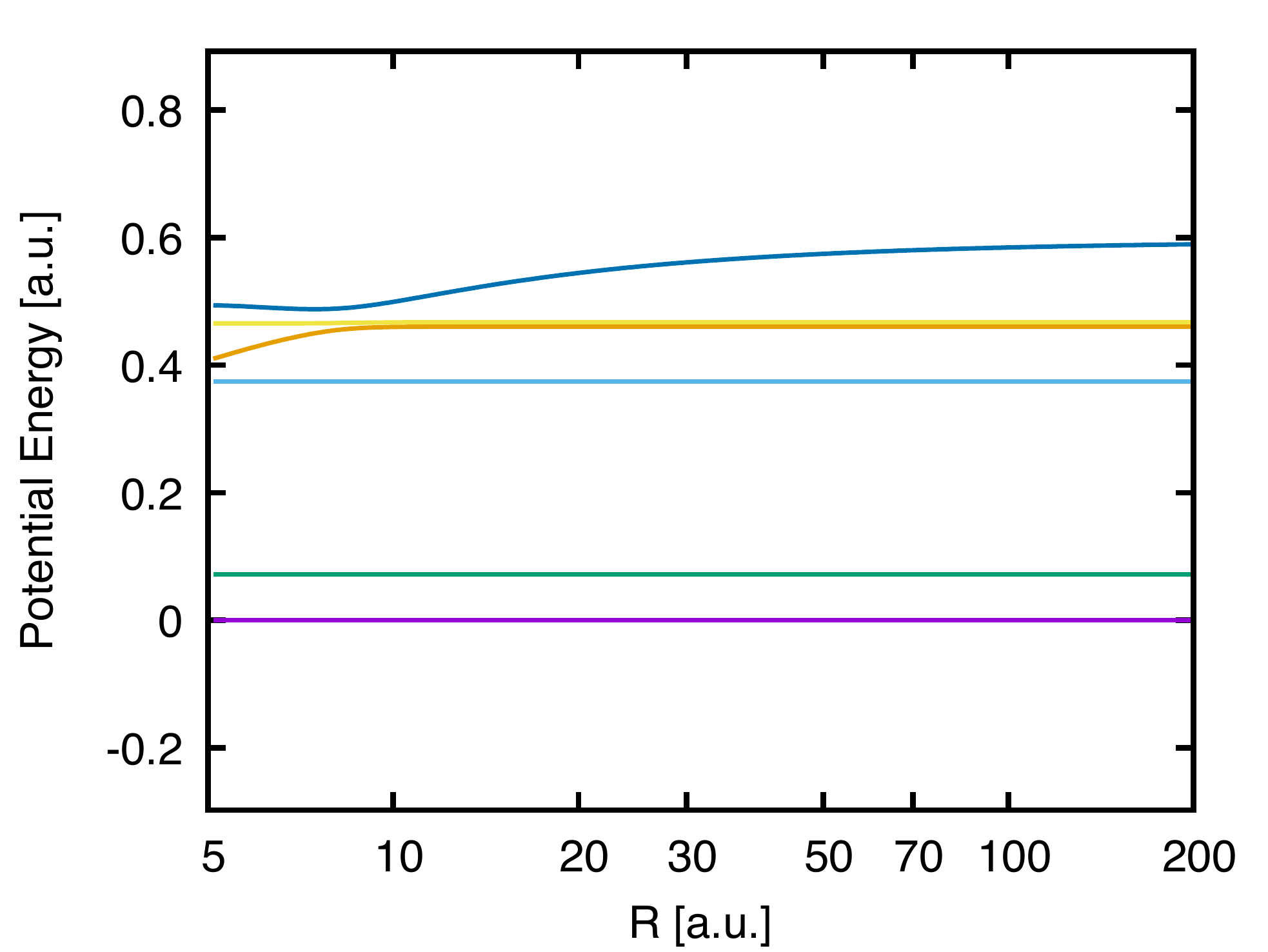}\put(25,17){$^2\Pi$, $\mathrm{O}^+(^2D^o)$ core}
\put(85,23){1}
\put(85,30){2}
\put(85,42){6}
\put(85,46){15}
\put(85,50){18}
\put(85,57){20}
\end{overpic}
\begin{overpic}[width=0.48\textwidth]{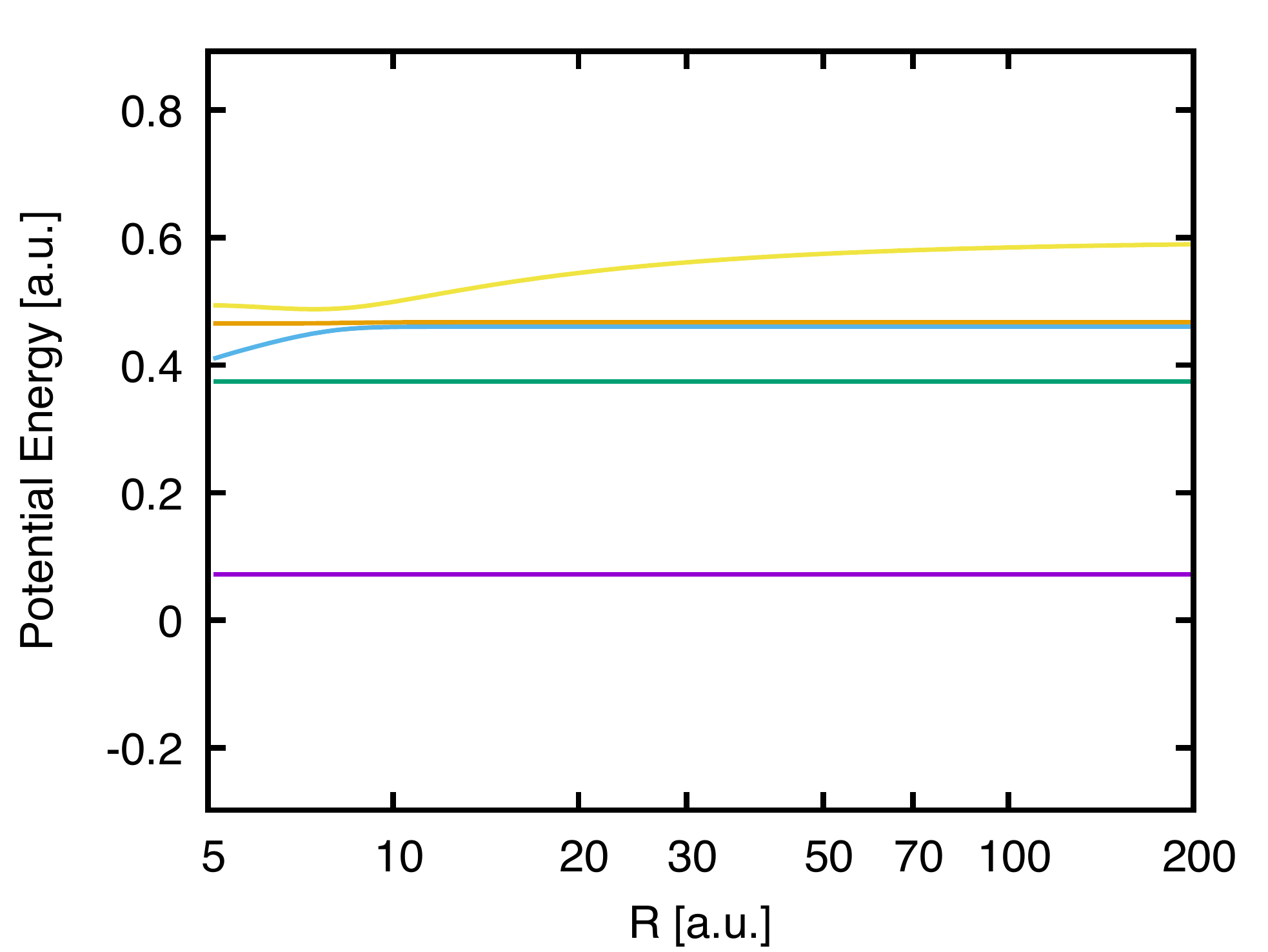}\put(25,17){$^2\Delta$, $\mathrm{O}^+(^2D^o)$ core}
\put(85,30){2}
\put(85,42){6}
\put(85,46){15}
\put(85,50){18}
\put(85,57){20}
\end{overpic}
\caption{Potential energies for O+H from the LCAO model for calculated symmetries and cores containing ionic crossings. The numbers on the right sides of the plots indicate the asymptotic states as given in Table~\ref{tab:syms}.}\label{fig:pots}
\end{figure*}

\begin{figure}
\centering
\begin{overpic}[width=0.48\textwidth]{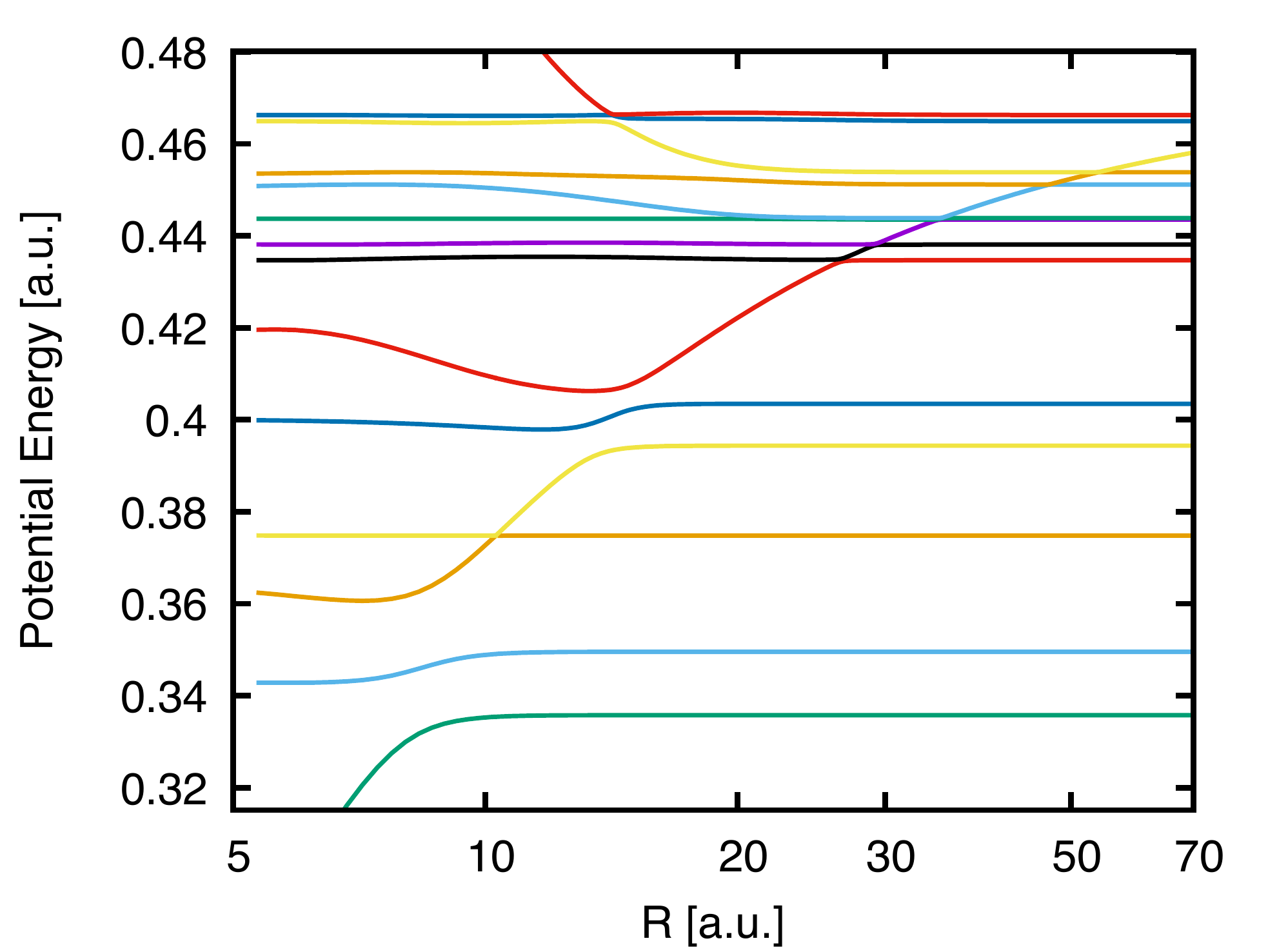}
\put(40,27){$^4\Sigma^-$, $\mathrm{O}^+(^4S^o)$ core}
\put(85,19){4}
\put(85,24){5}
\put(85,30){6}
\put(85,37){7}
\put(85,44){8}
\put(85,52){9}
\end{overpic}
\caption{Potential energies for O+H from the LCAO model, detailed view of $^4\Sigma^-$ with core $\mathrm{O}^+(^4S^o)$.  The numbers on the right side of the plot indicate the asymptotic states as given in Table~\ref{tab:syms}. }\label{fig:pots_zoom}
\end{figure}

\begin{figure}
\centering
\begin{overpic}[width=0.48\textwidth]{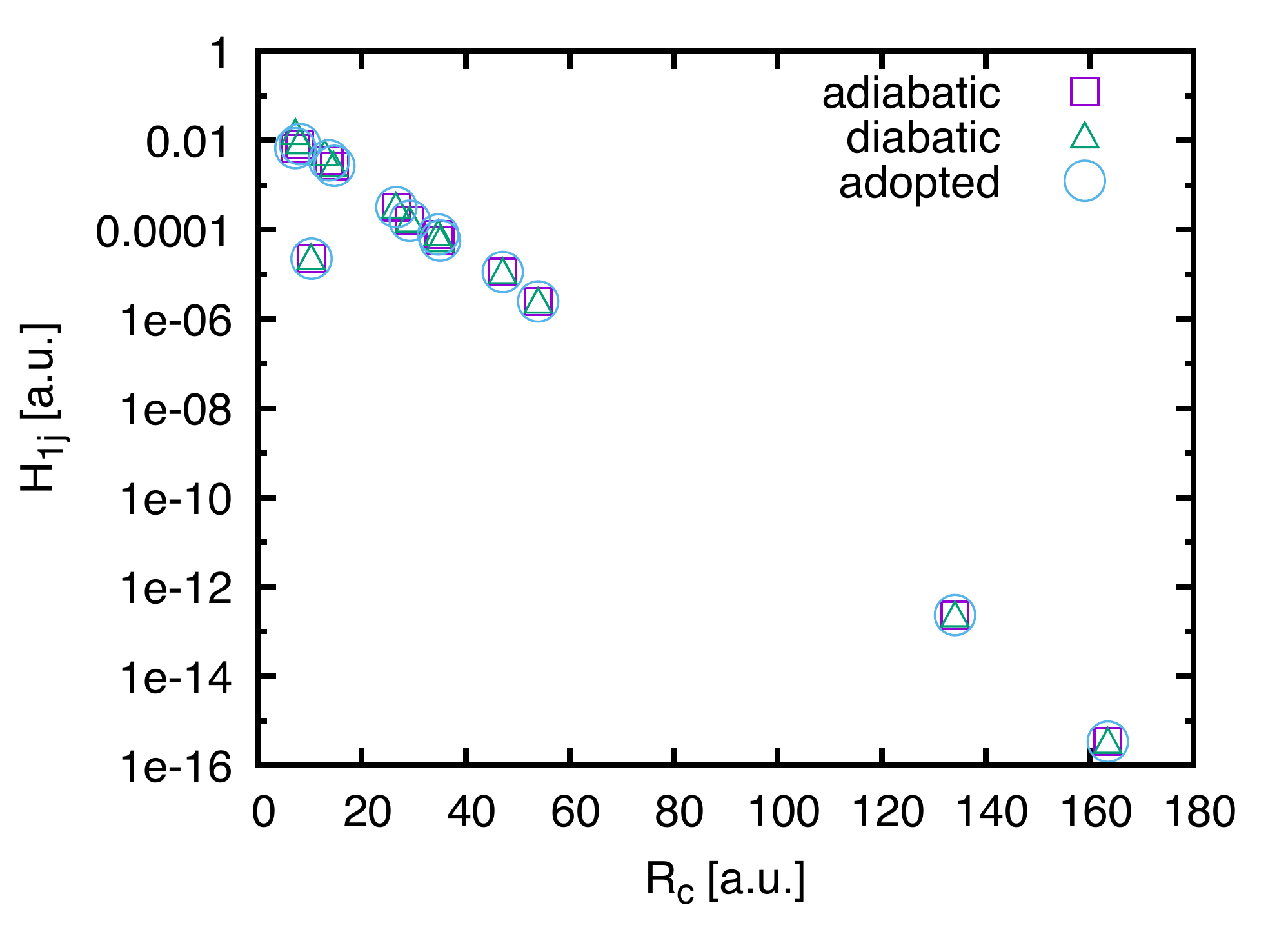}\put(25,27){$^4\Sigma^-$, $\mathrm{O}^+(^4S^o)$ core}\end{overpic}
\caption{Couplings $H_{1j}$ from the LCAO model plotted against the crossing distance $R_c$ for $^4\Sigma^-$ with core $\mathrm{O}^+(^4S^o)$. Results are shown for the adiabatic and diabatic models;  the final adopted values are also shown.  The point significantly lower than the general trend is the crossing with the covalent state $\mathrm{O}(2p^4.^3P)+\mathrm{H(n=2)}$. }\label{fig:lz}
\end{figure}

The resulting potentials for symmetries and cores where crossings occur are presented in Fig.~\ref{fig:pots}.  The first thing to be noted is that the ground covalent configuration $\mathrm{O}(2p^4.^3P,^1D,^1S)+\mathrm{H}$ has no crossings in the calculated range of internuclear distance, and thus no transitions in the present model.  Second, the crossings resulting from the $\mathrm{O}^+(^2D^o)$ core occur at quite short ranges and can be expected  not to give rise to processes with large cross sections and high rates.  The system of most interest is thus the $^4\Sigma^-$ symmetry resulting from the core corresponding to the $\mathrm{O}^+$ ground state $^4S^o$, as would be reasonably expected.  A more detailed view of the crossings in this system at intermediate internuclear distance ($\sim$30~$a_0$) is shown in Fig.~\ref{fig:pots_zoom}.  In particular, the crossing between the states labelled 6 and 7, shows a very small separation, which is due to the two-electron transition resulting from the fact that state 6 dissociates to $\mathrm{O}(2p^4.^3P)+\mathrm{H(n=2)}$.  In Fig.~\ref{fig:lz}, the couplings $H_{1j}$ for $^4\Sigma^-$ with core $\mathrm{O}^+(^4S^o)$ are plotted against the crossing distance $R_c$.  It is seen for this crossing that the coupling is between 2 and 3 orders of magnitude smaller than the typical couplings for one-electron transitions at similar internuclear distance.  It is also seen that the couplings derived from the adiabatic and diabatic models (see \S~II.B of B16) are in quite good agreement.  

\begin{figure}
\centering
\includegraphics[width=0.50\textwidth]{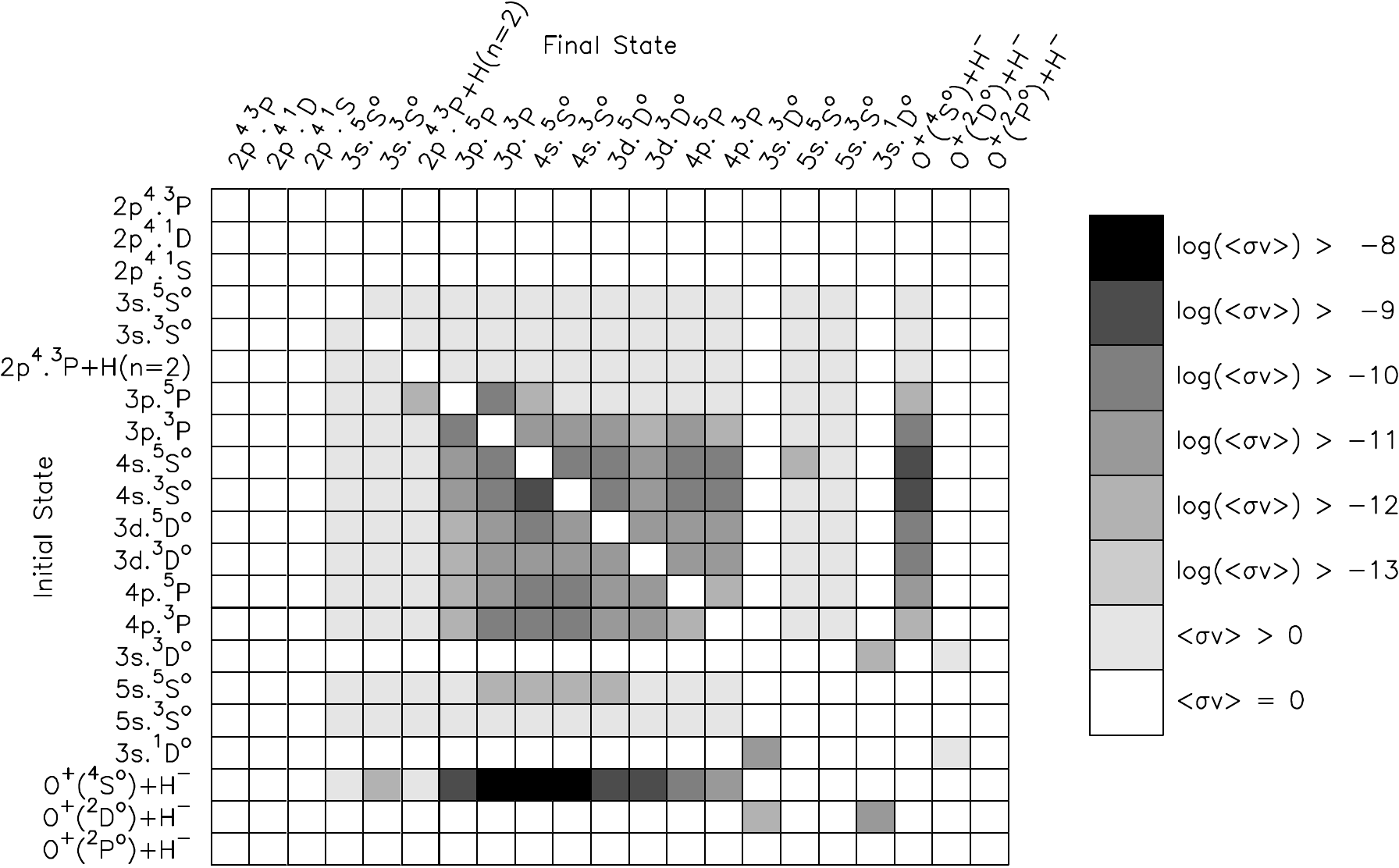}
\caption{Graphical representation of the rate coefficient matrix  $\langle \sigma \varv \rangle$ (in cm$^3$~s$^{-1}$) for inelastic O + H and O$^+$ + H$^-$ collisions at temperature $T = 6000$~K.  Results are from the LCAO asymptotic model.  The logarithms in the legend are to base 10. }\label{fig:rates_grid}
\end{figure}

\begin{figure}
\centering
\includegraphics[width=0.50\textwidth]{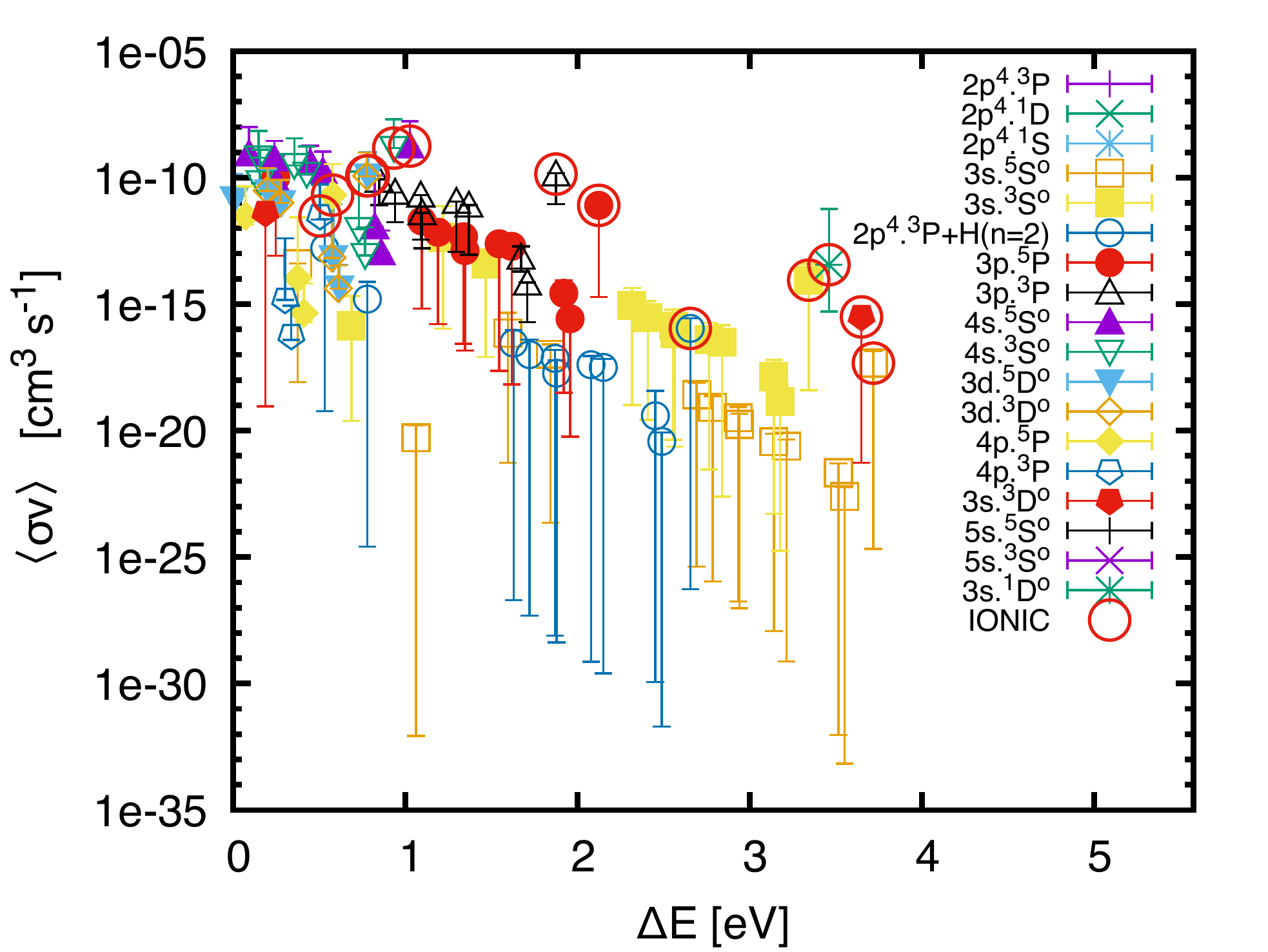}
\caption{Rate coefficients $\langle \sigma \varv \rangle$ for O+H collision processes at 6000~K, plotted against the asymptotic energy difference between initial and final molecular states, $\Delta E$.  The data are shown for endothermic processes, i.e. excitation and ion-pair production.  The   initial states of the transition are colour-coded, and the processes leading to a final ionic state (ion-pair production) are circled.  The points are from the LCAO model and the bars show the fluctuations.}\label{fig:rates}
\end{figure}

The LCAO model results for the rate coefficients, as well as the minimum and maximum values from alternate models, i.e. the `fluctuations' (see B16 and below), are published electronically at the CDS.  The rate coefficients $\langle \sigma \varv \rangle$ at 6000~K, typical for line forming regions in the solar atmosphere, from the calculations are presented in two different ways in Figs.~\ref{fig:rates_grid} and~\ref{fig:rates}; this temperature is assumed for the following discussion.  The fluctuations are shown in Fig.~\ref{fig:rates}, which plots the maximum and minimum rate coefficients calculated using the LCAO, SEMI-EMP, and LH-J models.  The LH-S model departs significantly from the other three models, leading to much larger fluctuations, in particular much lower results for processes with already low rates, and thus was omitted.  

The two figures demonstrate that, as has been found for other atoms, the largest endothermic processes from any given initial state are either charge transfer processes, specifically ion-pair production, or excitation processes to nearby states.  There are obvious reasons for this; charge transfer involves only one non-adiabatic transition, while excitation involves two, and nearby states have small thresholds.  In particular, it can be seen that ion-pair production from the $4s.^5S^o$ and $4s.^3S^o$ states gives the largest charge transfer and overall rate coefficients, and the transition between these two states gives the highest excitation rate coefficient, all three around $10^{-9}$ to $10^{-8}$~cm$^3/$s.  This result follows that seen in earlier work, where the first excited $S$-state often has the highest  rates, due to the crossings occurring at optimal internuclear distances, and $S$-states leading to high statistical weights \citep{Barklem2012}.  These highest rate coefficients show fluctuations typically of around one order of magnitude, and this may give some some indication of the uncertainty in the results for these processes where the ionic crossing mechanism can be expected to dominate.

    Processes involving low-lying states with configuration $3s$ have rate coefficients lower than $10^{-12}$~cm$^3/$s, due to rather small cross sections resulting from the fact that the crossings involving these states occur at rather short internuclear distance, $R<10$~$a_0$, with low transition probabilities, and thus lead naturally to small cross sections.  They also show large fluctuations due to strong (exponential) dependence of the transition probability on the coupling for these crossings with low transition probabilities.  Such low rate coefficients correspond to thermally averaged cross sections ($\langle \sigma \varv \rangle / \langle \varv \rangle$) of the order of $10^{-2}~a_0^2$ or less, and thus these processes could be dominated by contributions from other coupling mechanisms, namely radial couplings at short range and/or rotational and/or spin-orbit couplings.  The potential energy calculations of \cite{Easson1973} \citep[see also][]{Langhoff1982, vanDishoeck1983a} show a large number of states potentially interacting at short range, $R<3$~a$_0$, which could give rise to such mechanisms. 

Finally, processes involving the $\mathrm{O}(2p^4.^3P)+\mathrm{H(n=2)}$ configuration can been seen from Fig.~\ref{fig:rates} to have comparable or lower rate coefficients than other similar processes (e.g. those involving the nearby $3s$ and $3p$ states) and are typically low, less than $10^{-12}$~cm$^3/$s.  Thus, based on the estimate of the two-electron transition coupling used here, processes involving this configuration do not provide any high rates, at least via this coupling mechanism.

The only other calculations of inelastic O+H processes seem to be those for the process $\mathrm{O}(2p^4.^3P) + \mathrm{H}(1s) \rightleftarrows \mathrm{O}(2p^4.^1D) + \mathrm{H}(1s)$, which is important in a number of astrophysical environments.  \cite{Federman1983} performed Landau-Zener calculations based on curve crossings at short range between the relevant potential curves.  \cite{Krems2006} performed detailed quantum mechanical calculations using accurate quantum chemistry potentials and couplings \citep{vanDishoeck1983a, Parlant1999}.  The relevant couplings are due to spin-orbit and rotational (Coriolis) couplings \citep{Yarkony1992, Parlant1999}, and not the radial couplings due to the ionic crossing mechanism in the model used here.  Both \cite{Federman1983} and \cite{Krems2006} find rates of the order of $10^{-14}$~cm$^3/$s for the excitation process at 6000~K.  The calculations performed here lead to zero cross sections and rate coefficients because the ionic crossings are at extremely short range.

\section{Concluding remarks}

Excitation and charge transfer in low-energy O+H collisions has been studied using a method based on an asymptotic two-electron LCAO model of ionic-covalent interactions in the neutral atom-hydrogen-atom system, together with the multichannel Landau-Zener model (see B16).  The method has been extended to include configurations involving excited states of hydrogen, namely the $\mathrm{O}(2p^4.^3P)+\mathrm{H(n=2)}$ configuration, but this extension was found to  contribute only to processes with rather low rates when using an estimate for the two-electron transition coupling proposed by \cite{belyaev_atomic_2017}.  Rate coefficients are presented for temperatures in the range 1000--20000~K, and charge transfer and (de)excitation processes involving the first excited $S$-states, $4s.^5$S$^o$ and $4s.^3$S$^o$, are found to have the highest rates.  The fluctuations calculated with alternate models are around an order of magnitude, which may give some indication of the uncertainty in these rates.

Since the \ion{O}{i} triplet lines of interest in astrophysics correspond to $3s.^5S^o \rightarrow 3p.^5P$, rates involving these states may be of importance.  The work of \cite{Fabbian2009} shows the importance of the intersystem collisional coupling between the $3s.^3S^o$ and $3s.^5S^o$ states due to electron collisions, and \cite{amarsi_nonlte_2016}, when using the Drawin formula, has found the transition in the line due to hydrogen collisions to be of importance. The importance of these transitions seems to be borne out in preliminary statistical equilibrium calculations in stellar atmospheres (Amarsi, private communication).  Detailed testing of the sensitivity of the modelling to the data presented here will be carried out in the near future.  The estimates calculated here for processes $3s.^5S^o \rightarrow 3s.^3S^o$ and $3s.^5S^o \rightarrow 3p.^5P$ give very low rate coefficients, $\sim 4 \times 10^{-14}$~cm$^3/s$ and $\sim 7 \times 10^{-17}$~cm$^3/s$, respectively.  These low rates are uncertain since excitation processes involving these states are likely to have significant contributions from coupling mechanisms other than the ionic-covalent mechanism considered in the present model.  It should be noted that if the de-excitation transition $3s.^3S^o \rightarrow 3s.^5S^o$  had a collisional coupling with similar efficiency to that for $2p^4.^1D \rightarrow 2p^4.^3P$ from \cite{Krems2006}, the excitation rate coefficient would be of the order of $10^{-13}$ to $10^{-12}$~cm$^3/$s (noting the much smaller energy separation).   While there is no reason that the two transitions should behave in the same manner, this demonstrates the {possibility} for such couplings to provide significant contributions to the rate coefficients.  The possible deficiencies of the Landau-Zener model for such short-range crossings at near-threshold energies should also be borne in mind \citep[e.g.][]{Belyaev1999, Barklem2011}.  Modern quantum chemistry calculations including potentials and couplings and detailed scattering calculations for the low-lying states of OH, at least up to states dissociating to $\mathrm{O}(3p) + \mathrm{H}$ would be important to clarify this issue, and of potential importance in accurate modelling of the \ion{O}{i} triplet in stellar spectra, and thus in obtaining accurate oxygen abundances in late-type stars including the Sun. 

\begin{acknowledgements}
I thank Anish Amarsi for the comments on the draft, and for important preliminary feedback from astrophysical modelling.  I also thank Thibaut Launoy for the useful discussions that illuminated issues regarding the angular momentum coupling factors.  This work received financial support from the Swedish Research Council and the project grant ``The New Milky Way'' from the Knut and Alice Wallenberg Foundation.
\end{acknowledgements}

\bibliographystyle{aa} 
\bibliography{../../../MyLibrary.bib}

\end{document}